\documentclass[12pt]{amsart}
\usepackage{geometry} 
\usepackage{graphicx}
\usepackage{epstopdf}
\usepackage{natbib}
\usepackage{setspace}

\title[Emergent thresholds in genetic networks]{Emergent thresholds in genetic regulatory networks: Protein patterning in  \emph{Drosophila} morphogenesis}
\author{Rui Dil\~{a}o and Daniele Muraro}

\date{} 

\begin{document}
\maketitle

\begin{center}
Nonlinear Dynamics Group, Instituto Superior T\'ecnico,\\ Av. Rovisco Pais, 1049-001 Lisbon, Portugal.
\end{center} 

\

\begin{center}
rui@sd.ist.utl.pt, muraro@sd.ist.utl.pt 
\end{center}

\

\noindent \textbf{Keywords:} Systems Biology, Genetic Regulatory Networks

\begin{abstract}
We present a general methodology in order to build mathematical models of genetic regulatory networks. This
approach is based on the mass action law and on the Jacob and Monod operon model. The mathematical models are built symbolically by the \emph{Mathematica} software package  \emph{GeneticNetworks}. This package accepts as input the interaction graphs of the transcriptional activators and repressors  and, as output, gives the mathematical model in the form of a system of ordinary differential equations. All the relevant biological parameters are chosen automatically by the software.
Within this framework, we show that threshold effects in biology emerge from the catalytic properties of genes and its associated conservation laws. 
We apply this methodology to the segment patterning in \emph{Drosophila} early development and we calibrate and validate the genetic transcriptional network responsible for the patterning of the gap proteins Hunchback and Knirps, along the antero-posterior axis of the \emph{Drosophila} embryo. This shows that patterning at the  gap genes stage is a consequence of the relations between the transcriptional regulators and their initial conditions along the embryo. 
\end{abstract}

\section{Introduction}

A genetic regulatory network is an ensemble of interactions in a biological process involving  proteins, genes and mRNAs. The interactions between different proteins and genes can be  done by transcriptional activation and repression at the level of the genes, by protein-protein interactions, and by protein-mRNA interactions. 

A genetic regulatory networks is  described by a graph where vertices represent genes, proteins, enzymes or other chemical substances. The edges represent transformations, \textit{e. g.}, phosphorylation and dephosphorylation, or activation and inhibitory actions through transcription regulators. 

\begin{figure}[!htp]
\begin{center}
\includegraphics[width=3cm]{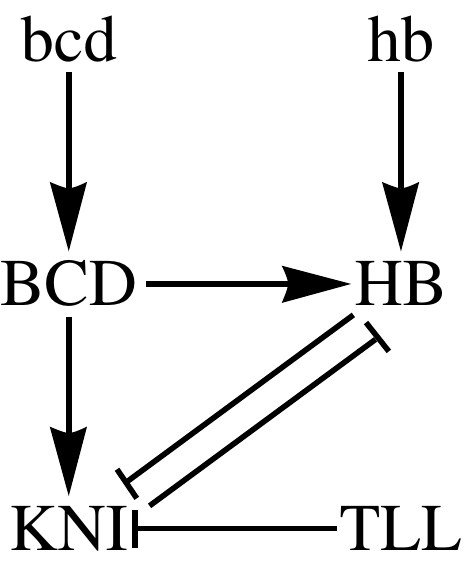}
\end{center}
\caption{Graph describing the genetic network  associated with the production of proteins Bicoid (BCD), Hunchback (HB), Knirps (KNI) and Tailless (TLL) in Drosophila early development. mRNAs \textit{hunchback} and \textit{bicoid} are represented by hb and bcd respectively. Arrows represent  activations and are listed in the set of ordered pairs $E_a$. Lines with perpendicular endings represent repressions and are listed in the set $E_r$.
}
\label{fig1}
\end{figure}

More precisely, a genetic regulatory networks is described  by a double graph $G=(V,E_a,E_r)$, where $V$ is the set of vertices or nodes of the graph and 
$E_a$ and $E_r$ are two sets of ordered pairs of vertices of the double graph. Each ordered pair of vertices defines the activation or the repression mechanism of the first node over the second. In classical graph theory, $G=(V,E_a)$ and $G=(V,E_r)$ are two  graphs with a common set of vertices. For example, in Figure~\ref{fig1}, we show the graph of a genetic network associated with the production of the proteins Bicoid (BCD), Hunchback (HB), Knirps (KNI) and Tailless (TLL) in \textit{Drosophila} early development, \citep{AD06}. In this example, we have $V=\{bcd, hb, BCD, HB, KNI, TLL\}$, $E_a=\{(bcd,BCD), (hb,HB),(BCD,HB),(BCD,KNI)\}$ and \break $E_r=\{(HB,KNI),(KNI,HB),(TLL,KNI)\}$. 

The graph of Figure~\ref{fig1} has a clear biological meaning. It expresses the fact that BCD is a transcriptional activator  of both HB and KNI,  HB and KNI proteins both repress each other, and TLL is a repressor of KNI.  The vertices  of the graph of Figure~\ref{fig1} can represent mRNAs, as in the case of hb and bcd, or proteins, as in the case of BCD and TLL, or genes and proteins simultaneously, as in the case of HB and KNI. 

Here we propose a set of rules in order to construct the model equations associated with a genetic regulatory networks described by a double graph $G=(V,E_a,E_b)$. This paper is an attempt to delineate a methodological approach for 
the construction of mathematical models of genetic regulation from the principles of chemical kinetics and chemical bound. In the literature, it is often found examples of mathematical models of biological systems described by different sets of equations  and characterized by different sets of parameters that are difficult to interpret and measure. Making qualitative predictions with these different models has a limited predictable value. For a review on the different approaches see, for example, \citep{klipp} and \cite{jong}.

From the point of view of predictability, it is important that 
mathematical model could be derived from first principles,
leading to a meaningful identification of the biological parameters, their calibration and validation with experimental data. 

In the construction of models for generic regulatory networks, we assume that models can be built with rate equations reflecting a mean field view of the stochastic random motion occurring at the molecular scale.  
This mean field approach, also called mass action law, is derived from the probabilistic collision laws occurring at the molecular scale. The models originated from this view are described by ordinary differential equations with polynomial vector fields, \citep{kampen}. 

One of the advantages of the mass action law approach is that the mean field rate equations have a direct microscopic interpretation, being associated with the collision mechanism that are in the origin of every reactive process. For model refinement, fluctuations can also be studied through the corresponding master equation.
From the experimental point of view,  microbiology techniques are strongly anchored in the mass action law or mean field approach. For example,  genome sequencing is only possible in the framework of the mean field approach.

For genetic regulatory networks described by graphs with a large number of vertices, and a complex structure of edges, the rate equations describing the evolution in time of concentrations are in general difficult to build, and are critically dependent on the assumptions done about the biological and chemical interactions involved.  During the development of these complex  models, it is often necessary to test different configurations changing the parameters and the initial conditions. Writing by hand all of these information is both time-consuming and error-prone.

In order to perform these tasks automatically, we have developed two \emph{Mathematica} software packages, \emph{Kinetics} and \emph{GeneticNetworks}, that execute the symbolic computations associated with the construction of the model equations for a genetic regulatory network. The result of the analysis is in symbolic form, and can be used in \emph{Mathematica} or in C for further processing and numerical integration. The software packages  \emph{GeneticNetworks} and \emph{Kinetics} are freely distributed from the site 
``https://sd.ist.utl.pt''. 

The \emph{Kinetics} package implements the mass action law in its polynomial exact form, computing symbolically the associated rate equations and  conservation laws. The
package \emph{GeneticNetworks} implements a particular model
for protein-gene regulation. This model for the protein-gene regulation is based on the operon model of Jacob and Monod \citep{JM61}, and its basic properties have been previously introduced in \cite{AD05}.

This paper is organized as follows.
In section \ref{mass} we briefly review the mass action law
of chemical kinetics and we introduce the collision graphs
associated with the mass action law. We derive the basic mass action rate equations. A special emphasis is done on mass action conservation laws, an important feature that is in the very foundations of threshold effects in biology. 
In other approaches, threshold do not result as emergent phenomena, but must be imposed through "ad-hoc" regulatory functions (see for example \citep[pp. 237]{klipp} or \citep{jong}).

In section \ref{genereg}, we describe the mechanism of genetic regulation based on the Jacob and Monod operon model, \citep{JM61}, and we introduce the modelling assumptions for
the construction of the mathematical models of genetic networks described by double graphs.

In section \ref{package}, we give an overview of the \emph{GeneticNetworks} software package. In
section \ref{applications}, we show three different applications of genetic regulatory networks. 
In the first application, we show, with a very simple example of autoregulation, that the conservation law constant is a bifurcation parameter for the regulation model, inducing a threshold effect in the production of proteins.  
In the second example, we give a genetic regulatory network inducing a localized spiky pattern.
In the third example, we analyze the experimental data associated with the KNI and HB inhibitory cross regulation in \textit{Drosophila} early development, described by the double graph of Figure~\ref{fig1}, and we calibrate this model with the experimental data. In the final section, we summarize the main biological conclusions of the paper.

\section{The mass action law framework of chemical kinetics}\label{mass}

In general, an ensemble of   chemical reactions is represented by the following collision diagram,
\begin{equation}
\nu_{i1} A_1  +  \cdots  + \nu_{im} A_m  \stackrel{r_i}{\longrightarrow} \mu_{i1} A_1  +  \cdots  + \mu_{im} A_m 
\label{eq1.1}
\end{equation}
where $i=1,\ldots , n$. The $A_j$, for $j=1,\ldots , m$, represent chemical substances, as for example, $A_j=\hbox{H}_2\hbox{O}$. The constants $\nu_{ij}$  and  $\mu_{ij}$ are the stoichiometric coefficients, in general, non-negative integers, and the constants $r_i$  are the rate constants.  If $\nu_{ij}=\mu_{ij}>0$, the corresponding substance $A_j$ is a catalyst and, if $\mu_{ij}>\nu_{ij}>0$,
$A_j$ is an autocatalyst. In the diagram (\ref{eq1.1}), there are $m$  chemical substances and $n$  rate constants or chemical reactions.
  
Under the hypothesis of  homogeneity of the solution where reactions occur, the mass action law asserts that the time evolution of the concentrations of the chemical substances is described by the system of ordinary differential equations,
\begin{equation}
\frac{{dA_j }}{{dt}} = \sum\limits_{i = 1}^n {r_i } (\mu_{ij}  - \nu_{ij} )A_1^{\nu_{i1} }  \cdots A_m^{\nu_{im} } 
\label{eq1.2}
\end{equation}
where $j=1,\ldots , m$, and we use the same symbol to represent both the chemical substance and its concentration.
The rate equations (\ref{eq1.2}) are derived under the following assumptions: (i) chemical reactions, when they occur, are due to elastic collisions between the reactants, (ii) homogeneity of the reacting substances in the solution, and (iii) thermal equilibrium of the solution.
All the kinetics aspects related with the dependence of the velocity of the reactions on the temperature or pressure are contained in the rate constants  $r_i$. 
For details see \cite{kampen}. 

At the atomic and molecular scale, chemical reactions between molecules can occur only if molecules collide or approach each other to small distances where bounding forces become meaningful. These chemical bounding forces are of electrical or quantum origin, and at distances larger than the mean free path they become less important when compared with the kinetics associated with the molecular motion. As chemical reactions only occur if  the chemical substances involved collide, the
vector fields associated with the right hand side of
(\ref{eq1.2}) are in general quadratic, representing binary collisions. Higher order polynomial vector fields are possible but, at the microscopic level, they are associated to triple or higher order collisions, a situations that occurs with a very low probability. Therefore, we will restrict our examples to models with two-body collisions.

The equations (\ref{eq1.2}) can also be written in the matrix form,
\begin{equation}
\frac{{dA}}{{dt}} = \Gamma \omega (A)
\label{eq1.3}
\end{equation}
where $\Gamma$  is a $n\times m$  matrix, $A^T  = (A_1 , \ldots ,A_m )$, and,
 \[
\omega(A) = 
\left( \begin{array}{c}
r_1 (\mu_{1j}  - \nu_{1j} )A_1^{\nu_{11} }\cdots A_m^{\nu_{1m} } \\
\vdots \\
r_n (\mu_{nj}  - \nu_{nj} )A_1^{\nu_{n1} } \cdots A_m^{\nu_{nm} }
\end{array} \right)
\]

In general, $n\not=m$, and the equations in system (\ref{eq1.2}) are not all independent. Let us denote by $r$  the rank of the matrix  $\Gamma$. The dimension of the null space of $\Gamma$  relates with its rank by, $ \dim (Null(\Gamma )) + r = m$   (number of columns of $\Gamma$). Let $v_1 , \ldots ,v_{m - r}$
 be a basis of the Null space of  $\Gamma$, then, 
$\Gamma v_k  = 0$, for $k=1,\ldots m-r$. So, by (\ref{eq1.3}), we have,
\begin{equation}
\frac{{dA}}{{dt}}.v_k  = \frac{d}{{dt}}(A.v_k ) = (\Gamma \omega ).v_k  = \omega .(\Gamma v_k ) = 0
\label{eq1.4}
\end{equation}
Hence, associated to the differential equations (\ref{eq1.2}), we have  the conservation laws,
\begin{equation}
A.v_k  = cons
\label{eq1.5}
\end{equation}
where,  $k=1,\ldots m-r$. 

The \emph{Mathematica} software package  \emph{Kinetics} calculates the rate equations (\ref{eq1.2}) describing the time evolution of the concentrations of the substances involved in the reactions described by the collision diagram (\ref{eq1.1}). The package calculates also the corresponding conservations laws (\ref{eq1.5}). 

The input of the package is the ensemble of chemical reactions, and the output of the package is the set of differential equations derived by the mass action law. The output can be later analyzed and studied by the analytical and numerical tools provided by Mathematica. In order to avoid long development times, the names of the rate constants are chosen automatically by the program.

The package \emph{Kinetics} has the usual  help commands, and we provide the \emph{Mathematica} notebook  \emph{KineticsTest.nb} with several self-explanatory examples and computations.

For example, let us now describe a simple protein production model with \emph{Kinetics}. The molecular biology dogma asserts that genes are the templates for protein production, and the standard  mechanism for protein production can be represented by the collision  diagrams,
\begin{equation}
\begin{array}{l}
Gene+Polymerase  \stackrel{r_1}{\longrightarrow} Gene+Polymerase+mRNA \\ 
mRNA  \stackrel{r_2}{\longrightarrow} mRNA+Protein\\
mRNA  \stackrel{r_3}{\longrightarrow}\\
Protein  \stackrel{r_4}{\longrightarrow}
\end{array} 
\label{eq1.6}
\end{equation}
Using the symbols $G$, $Pol$, $R$ and $P$ to represent gene, polymerase, mRNA and protein concentrations, respectively, the collision equations (\ref{eq1.6}) are the input for \emph{Kinetics}, with the syntax,
\begin{center} 
\texttt{input=$\{G+Pol\rightarrow G+Pol+mRNA,mRNA\rightarrow mRNA+P,mRNA\rightarrow W1, P\rightarrow W2\}$}
\end{center}
where $W1$ and $W2$ are waist products.

For the collision mechanism (\ref{eq1.6}), the rate equations  for the protein concentration, and  calculated by the package \emph{Kinetics}, are,
\begin{equation}\displaystyle
\begin{array}{l}
G'=0 \\ 
Pol'=0 \\ 
R'=r_1G.Pol-r_3R\\ 
P'=r_2 R -r_4P
\end{array} 
\label{eq1.7}
\end{equation}
and the rate constants have been chosen automatically by the  software package.
In this model, genes and mRNAs are catalysts, and these equations have the exact solutions,
\begin{equation}\displaystyle
\begin{array}{lcl}
G(t)&=&G(0) \\[2pt]
Pol(t)&=&Pol(0) \\[2pt]
R(t)&=&R(0)e^{-r_3t}+\frac{r_1G(0)Pol(0)}{r_3}(1-e^{-r_3t})\\ [2pt]
P(t)&=&G(0)Pol(0)\frac{r_1r_2}{r_3r_4}\\[2pt]
&+&\left(P(0)+\frac{R(0)r_2r_4-r_1r_2G(0)Pol(0)}{r_4(r_3-r_4)}\right)e^{-r_4t}\\[2pt]
&+&\frac{R(0)r_2r_3-r_1r_2G(0)Pol(0)}{r_3(r_4-r_3)}e^{-r_3t}
\end{array} 
\label{eq1.8}
\end{equation}

To simplify the model equations of protein production and maintaining the catalyst properties of genes, in the following, instead 
of the collision mechanism (\ref{eq1.6}), we use the simplified or reduced kinetic mechanism,
\begin{equation}
\begin{array}{l}
Gene  \stackrel{s_1}{\longrightarrow} Gene+Protein \\ 
Protein  \stackrel{s_2}{\longrightarrow} 
\end{array} 
\label{eq1.9}
\end{equation}
To the reactions (\ref{eq1.9}) correspond the rate equations,
\begin{equation}\displaystyle
\begin{array}{l}
G'=0 \\[5pt]
P'=s_1 G -s_2P
\end{array} 
\label{eq1.10}
\end{equation}

The rate equations (\ref{eq1.10}) have the solutions,
\begin{equation}\displaystyle
\begin{array}{l}
G(t)=G(0) \\[5pt]
P(t)=G(0)\frac{s_1}{s_2}+(P(0)-G(0)\frac{s_1}{s_2})
e^{-s_2t}
\end{array} 
\label{eq1.11}
\end{equation}

Comparing the protein solutions in (\ref{eq1.11}) and (\ref{eq1.8}), we conclude  that the steady state of protein of both models is unique and is proportional to the gene concentration. The proportionality constant is different for both models, depending on the rates of the reactions involved. The steady state $G(0)\frac{s_1}{s_2}$ of model (\ref{eq1.9}) has a direct biological meaning: $s_1$ is the rate of protein production and $s_2$ is the rate of protein degradation, and $G(0)$ is the initial gene concentration.

In the following, and in order to simplify the description of the transcriptional regulation of proteins, we will adopt the mechanism (\ref{eq1.9}) in order to describe protein production. 

In both rate equations  (\ref{eq1.10}) and (\ref{eq1.7}), the concentration of gene is constant along time, and therefore the gene concentration is a conservation law. In the following, we will show that the linear conservation laws of the form (\ref{eq1.5}), will have an important role in the determination of  steady states and in bifurcations associated with threshold effects.

\section{A mass action framework to describe genetic regulatory networks based on the protein-gene interaction}\label{genereg}

In the previous section, we have described a basic model 
model for the production of proteins, in the framework of the mass action law. Based on this framework, we  generalize this view in order to include the case of transcriptional regulation by proteins.

In order to keep general the approach presented here and to maintain the biological reality of model parameters, we make the following basic modeling assumptions:  

1) In order to describe quantitatively the protein production (concentration) within the molecular biology dogma, we only consider genes  and proteins. 
Intermediate substances in the regulatory mechanism like catalysts and mRNAs are not considered. A model of protein production has been presented and analyzed at the end of the previous section.

\begin{figure}[!htp]
\begin{center}
\includegraphics[width=4cm]{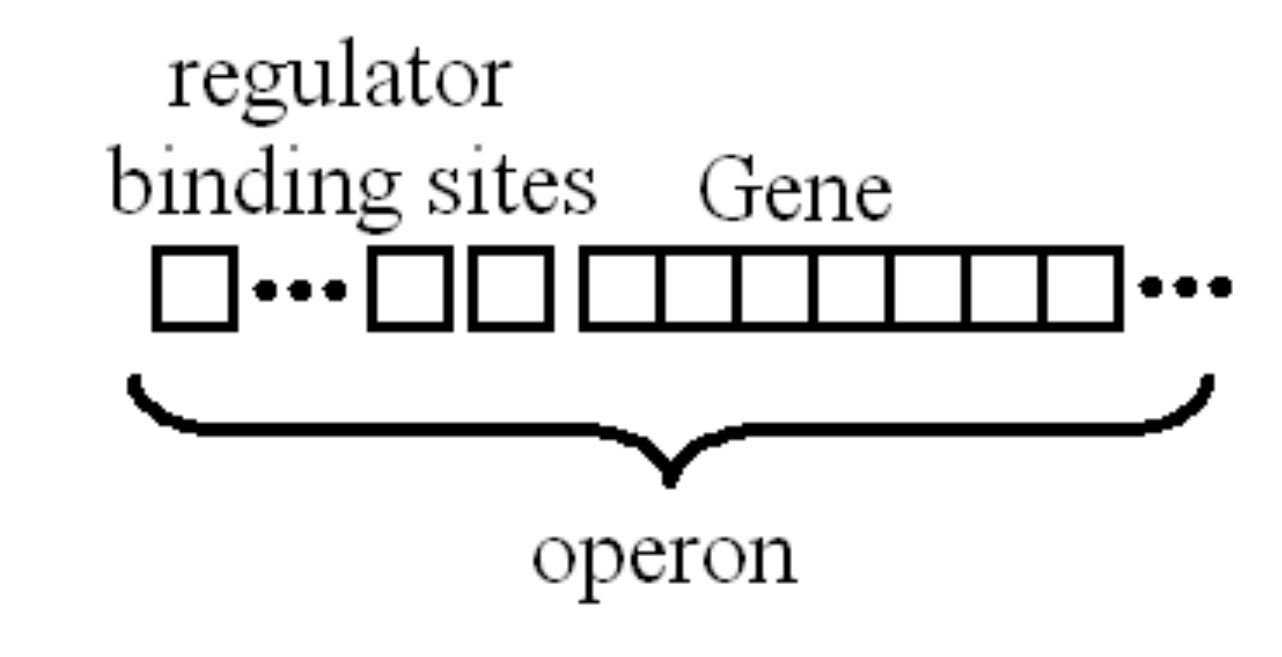}
\end{center}
\caption{Jacob and Monod operon model for the regulation of protein production. The transcription is regulated by the activators and the repressors binding to the binding sites of the gene. Figure adapted from \citep{AD05}.
}
\label{fig2}
\end{figure}

2) The regulation of protein production by the template gene is based on the Jacob and Monod operon model \citep{JM61}. 
Namely, every gene has associated a certain number of binding sites where transcription factors can bind --- activators or repressors, Figure~\ref{fig2}.  The regulation of activations and repressions occurs only through the binding sites. For a given double graph of interactions, the number of binding sites of a gene is determined by the number of edges that end up in the corresponding graph node.

3) Transcription factors are the proteins associated with the vertices that activate or inhibit the production of other proteins. The vertex of a graph represent a transcription factor only if it is the initial point of a edge of activation or inhibition. If a vertex has incoming and outcoming edges of any type, this vertex represents a protein and a gene  with several binding states.

4) Each  transcription factor has its own binding site in the gene strand, or each gene has only one binding site for all the regulators. Both cases are treated separately in the model.
We assume that when at least one activator is bound to a gene, the transcription is activated with a particular production rate for each combinatorial possibility of all the remaining binding sites. 
 
For example, in the double graph of the biological mechanism of Figure~\ref{fig1}, we have the following chemical substances, 
\begin{equation}\displaystyle
\begin{array}{ll}
\hbox{bcd, hb } & \hbox{mRNAs}\\ [2pt]
\hbox{BCD, HB, TLL} & \hbox{proteins}\\ [2pt]
\hbox{HB$^{0}_{0}$, HB$^{BCD}_{0}$, HB$^{0}_{KNI}$, HB$^{BCD}_{KNI}$}& \hbox{operons}\\ [2pt]
\hbox{KNI$^{0}_{0,0}$, KNI$^{BCD}_{0,0}$, KNI$^{0}_{HB,0}$, KNI$^{BCD}_{HB,0}$}& \hbox{operons}\\ [2pt]
\hbox{KNI$^{0}_{0,TLL}$, KNI$^{BCD}_{0,TLL}$, KNI$^{0}_{HB,TLL}$, KNI$^{BCD}_{HB,TLL}$}& \hbox{operons}
\end{array} 
\label{eq1.12}
\end{equation}
The description of the time evolution of protein concentrations of the mechanisms of Figure~\ref{fig1}  involves one rate equation for  each substance in (\ref{eq1.12}), except eventually for bcd and hb. As proteins are produced from a gene template, the symbol associated to each vertex of the graph represents a protein. The operon states are represented by the same symbol with superscripts and lowerscripts. The superscripts positions indicate the binding or unbinding  of transcriptional activators. The lowerscripts positions indicate the binding or unbinding  of transcriptional repressors. 

The model associated with a given double graph contains 
the rate equations for the proteins and the operons in its different states. We also assume by default that proteins always degrade and genes are autocatalytic substances that never degrade. The first assumption implies that protein concentrations remain bounded in time, and the second assumption implies the existence of a conservation law for the concentration of the operon states.

\section{The \emph{G\lowercase{enetic}N\lowercase{etworks}} software package}\label{package}

\emph{GeneticNetworks} is a software package that generates the rate equations for the concentrations of genes and  proteins in a regulatory network. The starting point is the double graph of activations and repressions, $G=(V,E_a,E_r)$. 
The inputs for \emph{GeneticNetworks} are two strings, \emph{activation} and \emph{repression}, that describe the transcriptional activations and repressions of proteins on genes. In the graph, the same symbol is used to denote both a gene and the corresponding produced protein. As we have seen in (\ref{eq1.12}), the set of variables for the regulation model is constructed with the vertex symbols and the associated operon states. 

\begin{figure}[!htp]
\begin{center}
\includegraphics[width=1.5cm]{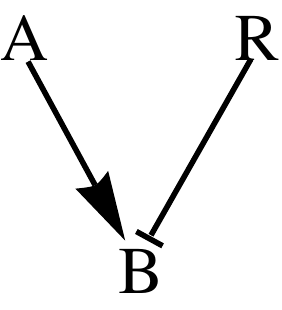}
\end{center}
\caption{Double graph associated with the input strings (\ref{eq2.1}) for the \emph{GeneticNetworks} software package.
}
\label{fig3}
\end{figure}

For example, using as input for \emph{GeneticNetworks} the interaction strings,
\begin{equation}
\begin{array}{lcl}
activation &=& \{ A \longrightarrow B \} \\
repression &=& \{ R \longrightarrow B \} 
\end{array}
\label{eq2.1}
\end{equation}
the double graph of the genetic network (\ref{eq2.1}) is shown in Figure~\ref{fig3}. In this case, the double graph $G=(V, E_a,E_r)$ is characterized by the sets, $V=\{A,B,R\}$,
$E_a=\{(A,B)\}$ and $E_r=\{(R,B)\}$.

In the interaction mechanism (\ref{eq2.1}), protein A activates gene B, and protein R represses gene B. Therefore, the variables of the mechanism (\ref{eq2.1}) are, 
\begin{equation}\displaystyle
\begin{array}{ll}
\hbox{A, B, R} & \hbox{proteins}\\
\hbox{B$_{0}^{0}$, B$_{0}^{A}$,B$_{R}^{0}$, B$_{R}^{A}$}& \hbox{operons}
\end{array} 
\label{eq2.2}
\end{equation}

The following functions are defined in the \emph{GeneticNetworks} package:
\begin{itemize}
\item \texttt{NetworkGraph}, \texttt{ManipulateGraph} 
\item \texttt{Reactions}, \texttt{ReactionsOneSite}, \texttt{ReactionGraph}
\item \texttt{SubstanceNames}, \texttt{SubstanceVariables}, \\ \texttt{SubstanceInitialConditions}
\item \texttt{ParameterNames}, \texttt{ParameterInput}
\item \texttt{Equations}
\item \texttt{ConservationLaws}
\end{itemize}
With these functions, we calculate the model equations associated with the input strings (\ref{eq2.1}), calculate automatically the number of variables of the model, define all the relevant parameters and  calculate the rate equations. For example, to the genetic regulatory network (\ref{eq2.1}), 
the \emph{GeneticNetworks} package associate the collision diagrams, 
\begin{equation}
\begin{array}{l}\displaystyle
A+B_0^0  \mathop{\buildrel\textstyle\longrightarrow\over
{\smash{\longleftarrow}\vphantom{_{\vbox to.5ex{\vss}}}}}_{b_{-a}}^{b_a} B_0^A
\\ \displaystyle
R+B_0^0  \mathop{\buildrel\textstyle\longrightarrow\over
{\smash{\longleftarrow}\vphantom{_{\vbox to.5ex{\vss}}}}}_{b_{-r}}^{b_r} B_R^0
\\ \displaystyle
R+B_0^A  \mathop{\buildrel\textstyle\longrightarrow\over
{\smash{\longleftarrow}\vphantom{_{\vbox to.5ex{\vss}}}}}_{b_{-r}}^{b_r} B_R^A
\\ \displaystyle
A+B_R^0  \mathop{\buildrel\textstyle\longrightarrow\over
{\smash{\longleftarrow}\vphantom{_{\vbox to.5ex{\vss}}}}}_{b_{-a}}^{b_a} B_R^A
\\ \displaystyle
B_0^A  \stackrel{\left(p_0^A\right)_B}{\longrightarrow} B_0^A+B \\ \displaystyle
B_R^A  \stackrel{\left(p_R^A\right)_B}{\longrightarrow} B_R^A+B \\ \displaystyle
B \stackrel{d_B}{\longrightarrow} 
\end{array} 
\label{eq2.3}
\end{equation}
To these collision diagrams, we have the mass action law rate equations,
\begin{equation}\displaystyle
\begin{array}{ccl}
 A'&=&-b_a A.B_R^0+b_{-a} B_R^A-b_a A.B_0^0+b_{-a}
   B_0^A\\
B'&=&\left(p_R^A\right)_B B_R^A+\left(p_0^A\right)_B B_0^A-d_B B\\
R'&=&-b_r R.B_0^A+b_{-r} B_R^A-b_r R. B_0^0+b_{-r} B_R^0\\
\left(B_0^0\right)'&=&-b_a A .B_0^0+b_{-a}
   B_0^A -b_r R 
   B_0^0 +b_{-r}
   B_R^0 \\
\left(B_0^A\right)'&=&b_a A.
   B_0^0 -b_{-a}
   B_0^A -b_r R.
   B_0^A +b_{-r}
   B_R^A \\
\left(B_R^0\right)' &=&-b_a A.
   B_R^0 +b_{-a}
   B_R^A +b_r R.
   B_0^0 -b_{-r}
   B_R^0 \\
\left(B_R^A\right)
   '&=&b_a A.
   B_R^0 -b_{-a}
   B_R^A +b_r R.
   B_0^A -b_{-r}
   B_R^A 
\end{array}
\label{eq2.4}
\end{equation}
and the  conservation law,
\begin{equation} 
\begin{array}{l}
B_0^0(t)+B_0^A(t)+B_R^0(t)+B_R^A(t)\\ 
=B_0^0(0)+B_0^A(0)+B_R^0(0)+B_R^A(0)
\end{array}
\label{eq2.5}
\end{equation}
From the conservation law (\ref{eq2.5}), we can eliminate one of the equations in (\ref{eq2.4}). In this genetic network, we have assumed that the  protein concentrations of $A$ and $R$ are constant along time.

The rate equations (\ref{eq2.4})
define a mass action law based model for the genetic regulatory network of Figure~\ref{fig3}.

In the implementation of \emph{GeneticNetworks}, we have two possible modeling choices. In one choice,  each different regulator has its own binding site, and the model diagrams (\ref{eq2.3}) have been constructed with this assumption. For the second choice, we consider that there is only one binding site in the operon  where all the regulators bind. In this   case, the collision diagrams associated with the genetic network (\ref{eq2.1}) and  calculated in the \emph{GeneticNetworks} package are,
\begin{equation}
\begin{array}{l}\displaystyle
A+B_0  \mathop{\buildrel\textstyle\longrightarrow\over
{\smash{\longleftarrow}\vphantom{_{\vbox to.5ex{\vss}}}}}_{b_{-a}}^{b_a} B_A
\\ \displaystyle
R+B_0  \mathop{\buildrel\textstyle\longrightarrow\over
{\smash{\longleftarrow}\vphantom{_{\vbox to.5ex{\vss}}}}}_{b_{-r}}^{b_r} B_R
\\ \displaystyle
B_A  \stackrel{p_B}{\longrightarrow} B_A+B \\ \displaystyle
B \stackrel{d_B}{\longrightarrow} 
\end{array} 
\label{eq2.6}
\end{equation}
By the mass action law, the \texttt{ReactionsOneSite} command leads to the rate equations,
\begin{equation}\displaystyle
\begin{array}{ccl}
 A'&=&b_{-a} B_A -b_a A.B_0 \\
B'&=&p_B B_A-d_B B\\
R' &=&b_{-r} B_R -b_r B_0.R\\
\left(B_0\right)' &=&-b_a  A . B_0 +b_{-a}B_A -b_r B_0.R+b_{-r} B_R\\
\left(B_A\right)'&=&b_a A. B_0-b_{-a} B_A\\
\left(B_R\right)'&=&b_r B_0. R-b_{-r}B_R 
\end{array}
\label{eq2.7}
\end{equation}
The equations (\ref{eq2.6}) have the conservation law, 
\begin{equation} 
B_0(t)+B_A(t)+B_R(t) 
=B_0(0)+B_A(0)+B_R(0)
\label{eq2.8}
\end{equation}

The two models (\ref{eq2.3}) and (\ref{eq2.6}) for the genetic network (\ref{eq2.1}) are different and these two choices are implemented in the \emph{GeneticNetworks} package. For the dynamical analysis of a particular case of the distinction between the two
models (\ref{eq2.3}) and (\ref{eq2.6}), see \citep{AD05}. Below, we will show with a specific example that these two different regulation choices lead to qualitatively and quantitatively similar results.

\section{Applications}\label{applications}

\subsection{An emerging threshold in the dynamics of a self-activating protein}\label{example1}

\begin{figure}[!htp]
\begin{center}
\includegraphics[width=1.5cm]{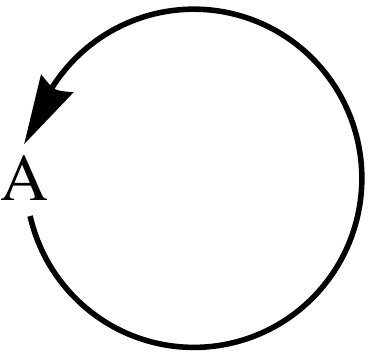}
\end{center}
\caption{Regulation graph describing a self-activating protein.
}
\label{fig4}
\end{figure}

As an application of the rules describing a genetic regulatory network just introduced, we discuss now the basic role of the conservation laws in the occurrence of threshold effects in regulation mechanisms. We study the case of a self-activating protein, where  the produced protein activates its own production, Figure~\ref{fig4}.

The simplest self-activating genetic network is described by the input tables, 
\begin{equation*}
\begin{array}{lcl}\displaystyle
activation&=&\{A \rightarrow A\} \\
repression&=&\{ \}\, .
\end{array}
\end{equation*}
The reactions and the parameters involved in this activation can be obtained by the \emph{GeneticNetworks} command \texttt{Reactions}, followed by the command \texttt{ReactionGraph},
\begin{equation}
\begin{array}{l}\displaystyle
A^0+A  \mathop{\buildrel\textstyle\longrightarrow\over
{\smash{\longleftarrow}\vphantom{_{\vbox to.5ex{\vss}}}}}_{a_{-a}}^{a_a} A^A
\\ \displaystyle
A^A \stackrel{p_A}{\longrightarrow} A^A+A \\ \displaystyle
A \stackrel{d_B}{\longrightarrow} 
\end{array} 
\label{reactions1}
\end{equation}
where $A^0$ and $A^A$ are operon states and $A$ is the corresponding protein.

With the command \texttt{Equations}, we get the rate equations,
\begin{equation}
\begin{array}{lcl}\displaystyle
A' & = & a_a A .A^0 + a_{-a}A^A +p_A A^A -d_A A\\
(A^0)' & = & - a_a A. A^0 + a_{-a} A^A  \\
(A^A)'& = & a_a A .A^0 - a_{-a}A^A  
\end{array}
\label{equations1}
\end{equation}
Finally, with the command \texttt{ConservationLaws}, we find,
\begin{equation}
A^0 (t) + A^A (t) =A^0 (0) + A^A (0)= c
\label{conservation1}
\end{equation}
where $c$ is a constant.

Introducing (\ref{conservation1}) into (\ref{equations1}), the independent set of rate equations describing the process (\ref{reactions1}) is,
\begin{equation}
\begin{array}{lcl}\displaystyle
A'&=& a_a A .A^0 + a_{-a}( c-A^0 ) +p_A ( c-A^0 ) -d_A A \\ \displaystyle 
(A^0)'  &=& - a_a A .A^0 + a_{-a} ( c-A^0 ) 
\end{array}  
\label{indipendentEquations1}
\end{equation}

We analyze now the steady state and the phase space structure of the solutions of equations (\ref{indipendentEquations1}).
Equations (\ref{indipendentEquations1}) have two steady states with coordinates,
$$
(A_{(1)}^{0},A_{(1)})=(c,0)
$$
and,
$$
(A_{(2)}^{0},A_{(2)})=\bigg(\frac{d_A a_{-a}}{p_A a_a},\frac{c p_A  a_a - d_A a_{-a}}{d_A a_a}\bigg)
$$
As the coordinate of the two fixed points are dependent of $c$, by (\ref{conservation1}), the steady state coordinates are dependent of the initial concentrations of the operon. 

Let $J_{(i)}$, $i=1,2$ be the Jacobian of equation (\ref{indipendentEquations1}) evaluated at the fixed points. As,
\[
\det{J_{(1)}}=-\det{J_{(2)}}
\]
and,
\[
\det{J_{(1)}}<0 \Longleftrightarrow a_{-a} d_A < c a_a  p_A
\]
then, we have, 
\begin{eqnarray}\label{bifurcation}
\textrm{if}\ a_{-a} d_A < c a_a p_A &  & \left\{ \begin{array}{l}
(A_{(1)}^{0},A_{(1)})\ \textrm{is of saddle type}\\
(A_{(2)}^{0},A_{(2)})\ \textrm{is a stable node}
\end{array} \right. \label{bifurcation1}\\
\textrm{if}\ a_{-a} d_A > c a_a p_A &  & \left\{ \begin{array}{l}
(A_{(1)}^{0},A_{(1)})\ \textrm{is a stable node}\\
(A_{(2)}^{0},A_{(2)})\ \textrm{is of saddle type}
\end{array} \right. \label{bifurcation2}
\end{eqnarray}
As, for $c<a_{-a}d_A / a_a p_A$, $A_{(2)}$ is negative, 
the protein concentration at the steady state of the rate equations (\ref{indipendentEquations1}) is zero ($A_{(1)}=0$). For $c>a_{-a}d_A / a_a p_A$, the  protein concentration at equilibrium is $A_{(2)}=\frac{c p_A  a_a - d_A a_{-a}}{d_A a_a}$.
Therefore,
the conservation law (\ref{conservation1}) tune a bifurcation for $c=a_{-a}d_A / a_a p_A$ (transcritical bifurcation), implying 
the existence of a threshold effect tuned by the conservation law parameter.

\begin{figure}[!htp]
\begin{center}
\includegraphics[width=5cm]{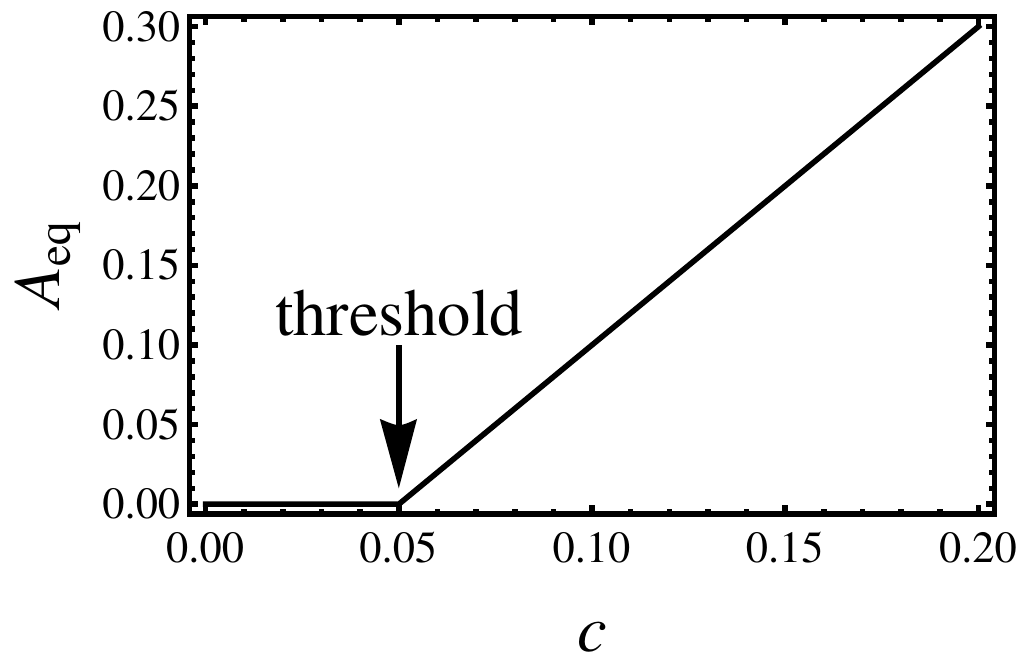}
\end{center}
\caption{Dependence of the protein steady state on the total concentration of the gene. Below the bifurcation or threshold $c=a_{-a}d_A / a_a p_A$, the equilibrium value of protein concentration $A_{eq}$ is zero, while above it takes the value $A_{eq}=( p_A c a_a - d_A a_{-a})/(d_A a_a)$. The parameters are: $a_a =1.0$, $a_{-a}=0.1$, $p_A =0.2$ and $d_A =0.1$. 
}
\label{fig5}
\end{figure}

In Figure~\ref{fig5}, we show the dependence of the protein steady state on the total concentration of the gene.

\subsection{Spatial distribution and steady states}\label{example2}

We have focused on genetic regulatory models without specifying a spatial localization. In a  genetic network describing some biological process, the initial concentration of  proteins can significantly vary in space, across tissues. This is specifically true in developmental processes, where often proteins show a spatial distribution with very sharp slopes that play a basic role in the establishment of body plans of organisms. A well known case is the \textit{Drosophila} segmentation, where variations on protein concentrations across the embryo induces protein patterning, \citep{Driever, Nuss}. One of such genetic regulatory networks is the one represented in Figure~\ref{fig1}, \citep{AD06}.

\begin{figure}[!htp]
\begin{center}
\includegraphics[width=1.5cm]{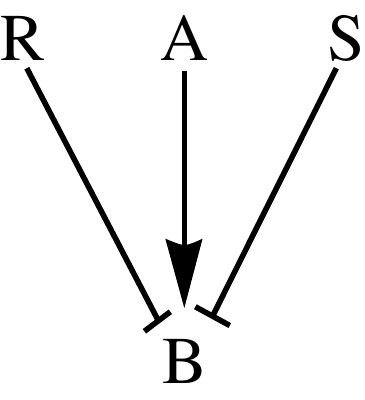}
\end{center}
\caption{Genetic regulatory network for the production of protein $B$ with one activator $A$ and two repressors $R$ and $S$.
}
\label{fig6}
\end{figure}

To show that patterning can be explained by the non homogeneity of initial conditions of regulators across tissues, we analyze a genetic regulatory network for the production of a protein $B$, regulated by one activator $A$ and two repressor proteins $R$ and $S$, Figure~\ref{fig6}. To simplify our analysis, we take the competitive case, where the activator and the repressors bind to the same binding site of the operon of protein $B$. 

To simplify further, we assume that the spatial distribution of the proteins $A$, $R$ and $S$ are constants. Under these conditions and with the regulation model developed in the package \emph{GeneticNetworks}, we obtain the system of linear rate equations,
\begin{equation}
\begin{array}{lcl}\displaystyle
B'  & = & p_B B_A - d_B B \\
(B_A)' & = & b_a A B_0 - b_{-a}B_A  \\
(B_R)'& = & b_r R B_0 - b_{-r}B_R  \\
(B_S)'& = & b_s S B_0 - b_{-s}B_S  
\end{array}
\label{equations3}
\end{equation}
where 
$B=B(x,t)$, $B_0=B_0 (x,t)$, $B_A=B_A (x,t)$, 
$B_R=B_R (x,t)$, $B_S=B_S (x,t)$,
$A=A(x)$, $R=R(x)$ and $S=S(x)$. These concentration variables are considered to be defined in a spatial one-dimensional bounded region of the real line ($x\in I\subseteq \mathbb{R}$). The following conservation law holds, 
$$
B_0(x,t) + B_A(x,t) + B_R(x,t) + B_S (x,t)= c(x)
$$
where $c(x)$ is a constant, depending eventually of the spatial independent coordinate $x$. 
The system of rate equations (\ref{equations3}) has one steady state with coordinates,
\begin{equation*}
\begin{array}{lcl}\displaystyle
\bar{B} = \frac{p_B \bar{B}_A}{d_B} &\, ,\, & \displaystyle\bar{B}_A = \frac{c A b_a b_{-r}b_{-s}}{D}\\[6pt] \displaystyle
\bar{B}_R = \frac{c R b_{-a} b_r b_{-s}}{D}&\, ,\, & \displaystyle \bar{B}_S = \frac{c S b_{-a} b_{-r}b_s }{D} 
\end{array}
\end{equation*}
where,
\[
D = b_{-a} b_{-r} b_{-s}+A b_a b_{-r} b_{-s}+R b_{-a} b_r b_{-s}+S b_{-a} b_{-r}b_s
\]
Choosing $b_a = b_r = b_s$, and $b_{-a}=b_{-r}=b_{-s}$, the steady state concentration of the protein $B$, is,
\begin{equation}
\bar{B}(x)=\frac{c(x) A(x) p_B}{(1+A(x)+R(x)+S(x))d_B}
\label{SteadyEq}
\end{equation}

\begin{figure}[!htp]
\begin{center}
\includegraphics[width=4.8cm]{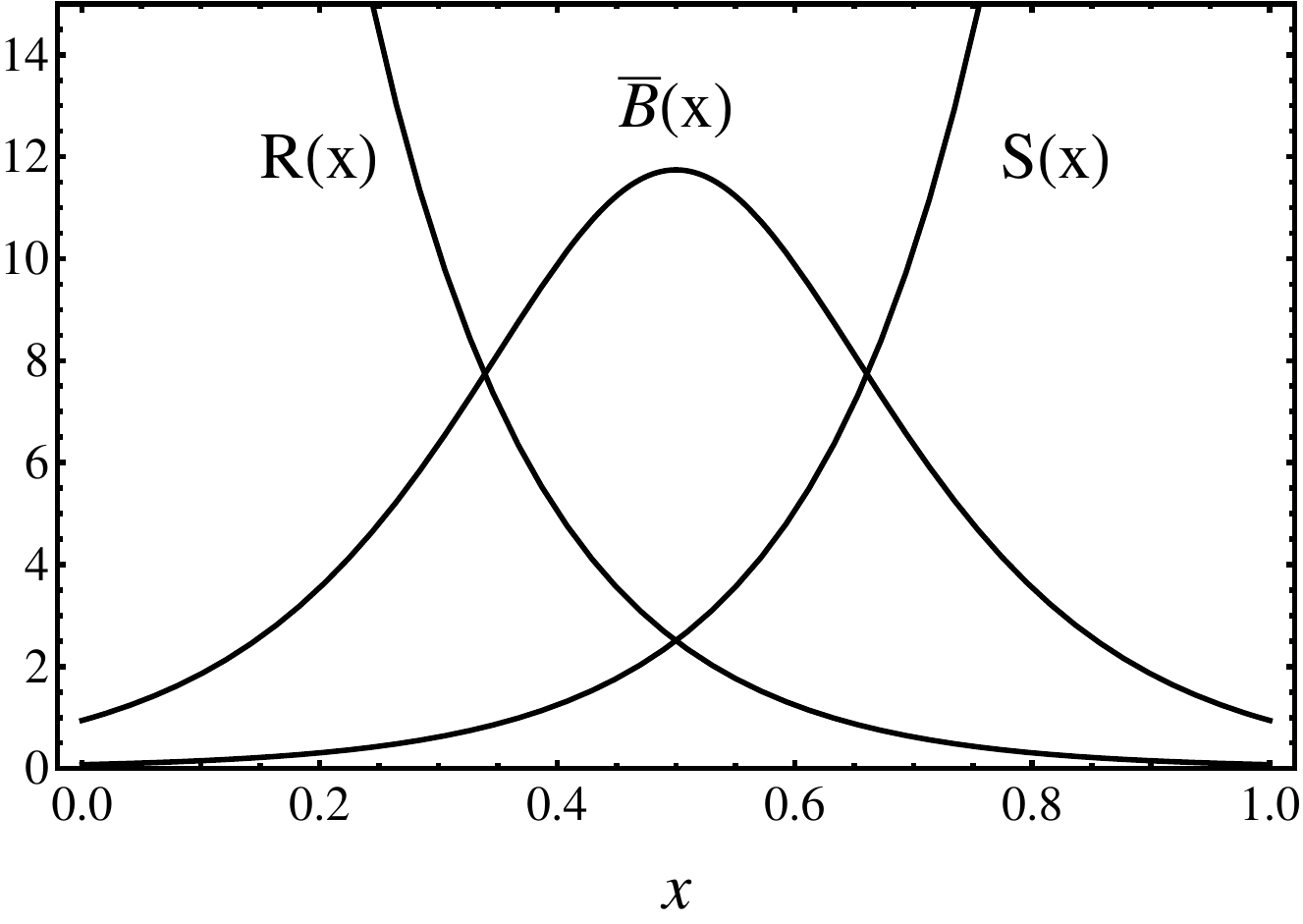}
\end{center}
\caption{Steady states of proteins $B$, $R$ and $S$ for the genetic regulatory network of Figure~\ref{fig6}. The steady state of protein $B$ shows a spiky profile, resulting from the inhibitory action of proteins $R$ and $S$. In this model, we have considered that the concentrations of $R$ and $S$ are constant in time and non-homogeneous in space. The activator protein $A$ has been considered constant along the spatial region. 
}
\label{fig7}
\end{figure}

In Figure~\ref{fig7}, we show the steady state concentration (\ref{SteadyEq}) of protein $B$, as a function of a spatial coordinate, $x\in [0,1]$. We have 
considered the initial distributions $A(x)=0.8$, $R (x)=83 e^{-7x}$, $S (x)=83 e^{-7 (1-x)}$, $c(x)=100$, and the parameter value $p_B/d_B=1$. For this case, the steady distribution of protein $B$ is spiky, due to the inhibitory regulation of the repressors.
We have analyzed the same genetic network of Figure~\ref{fig6} with a model with different binding sites for each regulator. The final result is similar with the one  shown in Figure~\ref{fig7}.
This shows that the two model approaches in \emph{GeneticNetworks}, with one binding site and with several binding sites in the operon, give similar qualitative results. When, the calibration and validation of models is not a problem, we can use the simplest one binding site regulation model in order to describe a given genetic regulatory network. The one-regulator site framework leads to simpler mathematical models. 

The  solution  (\ref{SteadyEq}) shows that steady states can depend on the initial conditions of the regulators ($A(x)$, $R(x)$ and $S(x)$) and on the initial conditions of the catalytic agents ($c(x)$), showing that spatial patterning can be a direct consequence of the non homogeneity of initial conditions. 

In this model, we have considered that the concentrations of
$A$, $R$ and $S$  are constants, implying that the model equations (\ref{equations3}) describe a system where  $A$, $R$ and $S$ have a fast recovery time.  This situation only occurs in thermodynamically open systems, as is the case of biological systems.

\subsection{Cross-regulation in \emph{Drosophila}}

\begin{figure}[!htp]
\begin{center}
\includegraphics[width=7.4cm]{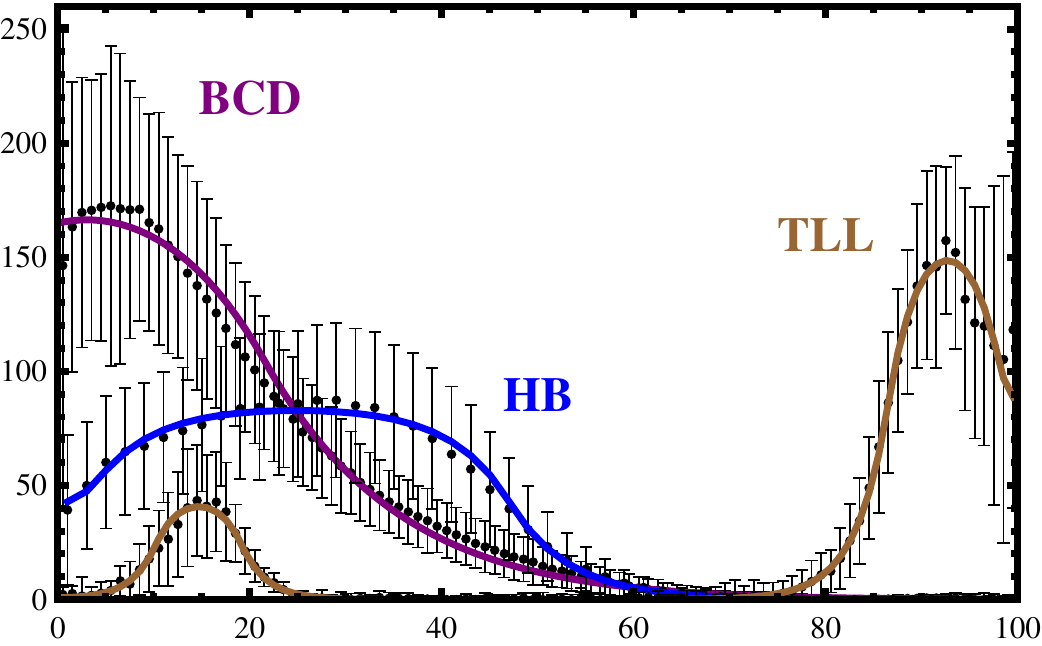}
\end{center}
\caption{Concentration of protein Hunchback (HB) at the end of cleavage cycle 13, and of Bicoid (BCD) and Tailless (TLL) proteins at the cleavage cycle 14, along the antero-posterior axis of the \emph{Drosophila} embryo. 
The embryo length has been scaled from $0$ to $100$, and the vertical axis units are proportional to protein concentration.
The continuous curves represent the mean distribution of the concentration of proteins calculated for the data of $954$ embryos. The data has been taken from the FlyEx database 
\citep[http://flyex.ams.sunysb.edu/flyex/]{Poustelnikova,Pisarev}. 
}
\label{fig8}
\end{figure}

In \textit{Drosophila} early development, some proteins as Bicoid (BCD) and Hunchback (HB) are translated from mRNA of maternal origin. Early in the first developmental stages of \textit{Drosophila}, at cleavage stage 13, BCD and HB proteins form a stable gradient along the antero-posterior axis of the \textit{Drosophila} embryo, Figure~\ref{fig8}. This stable gradient has it origin in a diffusion process occurring in the syncytial blastoderm of the embryo, \citep{DM09}. At a latter stage, in the 14th cleavage stage, other proteins as Knirps (KNI) show segments characterized by spiky concentration patterns along the antero-posterior axis of the embryo. 

We  show now  that the  patterning HB and KNI proteins as observed at late cleavage stage 14 of the embryo of \textit{Drosophila} is due to the non homogeneity of protein distribution at an early developmental stage. For that,
we have taken the genetic regulatory network of Figure~\ref{fig1}, describing the genetic regulation of HB and KNI, and we have  used the package \emph{GeneticNetworks} to describe the production of proteins Hunchback and Knirps during the cleavage cycles 14 of the developing embryo of \textit{Drosophila}. 

At the end of cleavage cycle 13 and beginning of the 14th,  BCD, HB and TLL proteins have a non homogeneous distribution along the embryo, Figure \ref{fig8}.

\begin{figure}[!htp]
\begin{center}
\includegraphics[width=7.4cm]{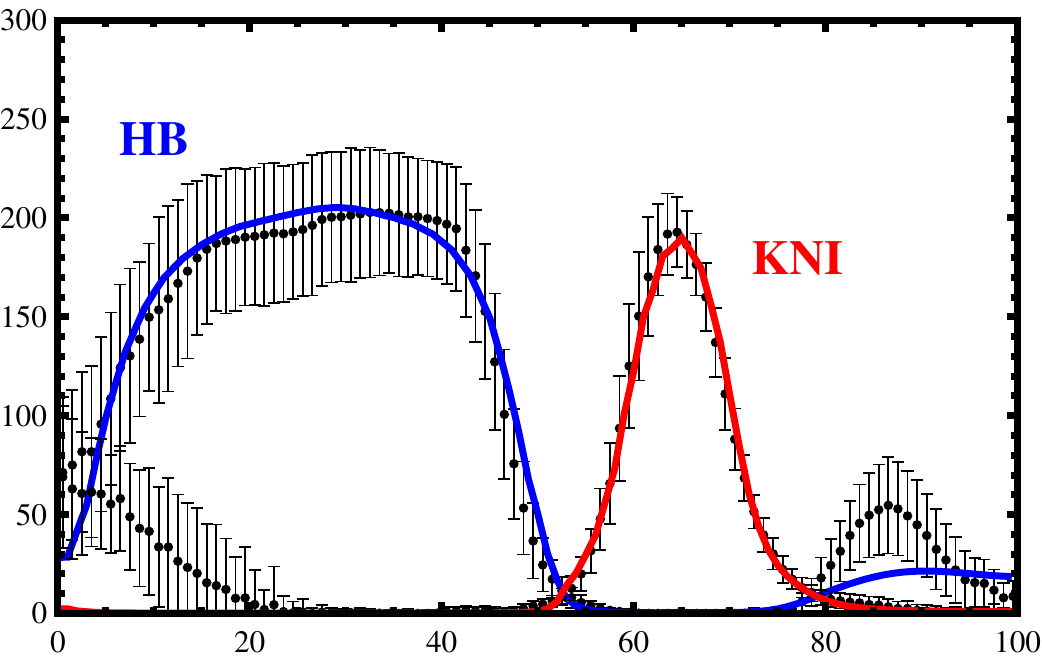}
\end{center}
\caption{The dots with error bars are the mean experimental profiles of the proteins HB and KNI at cleavage cycle 14, for several  \textit{Drosophila} embryos along the antero-posterior axis. 
The embryo length has been scaled from $0$ to $100$, and the vertical axis units are proportional to protein concentration. The lines show the prediction of the model profile for proteins HB and KNI, obtained with the software package \emph{GeneticNetworks}.
The mean distribution of the concentration of proteins has been calculated for the data of $954$ embryos, taken from the FlyEx database \citep[http://flyex.ams.sunysb.edu/flyex/]{Poustelnikova,Pisarev}. }
\label{fig9}
\end{figure}

Hunchback and Knirps proteins are both activated by the maternally produced protein Bicoid, Figure \ref{fig1}, and they  mutually repress each other.
The protein Knirps is also repressed by the protein  Tailless of maternal origin. Therefore, the genetic network model obtained with the package \emph{GeneticNetworks} should lead to the experimental profiles of HB and KNI, as observed at cleavage cycle 14, Figure~\ref{fig9}.

In  Figure~\ref{fig9}, we show the experimental profiles of HB and KNI at cleavage cycle 14, as well as a fit of the experimental data obtained with a model built with the software package \emph{GeneticNetworks}. The model equations have $22$ free parameters. To fit the experimental data with the mathematical model, we have assumed that the initial protein concentrations of BCD and TLL are constant over time, and we have assumed the option of several binding sites per operon. To find the numerical values of the 
model parameters in order to calibrate the model, we have used an optimization techniques based on a genetic algorithm, \citep{Dilao, Dilao2}. 
We have considered that, in the ordinary differential equations of the model, the time is also a free parameter. The protein profiles shown in Figure~\ref{fig9} are out of steady state pattern, obtained for the integration time $t=805$~s.

The  fitted curves in Figure~\ref{fig9} shows a very good agreement with the experimental data. HB fitts well in the embryo length range $[5\%,80\%]$, and KNI fittes well in the embryo length range $[20\%,100\%]$. The quality of the fits has been measured by the penalized chi square test. For the two fits in Figure~\ref{fig9}, we have obtained the reduced chi square values $\chi^2_{HB}=0.57$ and $\chi^2_{KNI}=0.68$. This  validates the genetic regulatory network of Figure~\ref{fig1}. This suggest that the anterior and posterior regions of the embryo are under the control of additional regulators.

\section{Conclusion}

We have presented a software tool to build mathematical models of genetic regulatory systems. The input of the package is the  graph  containing the list of transcriptional activators and repressors of the network.  The software implements an approach based on the mass action law, on the operon  model for regulation, and we have assumed in general that genes are always catalytic substances presented in any genetically controlled biological process.   

Within this approach, the usual threshold concept in biology emerges as a bifurcation phenomenon of the model equations. These bifurcations are in general tuned by the conservation law constants of the model equations. Therefore, within this framework, thresholds are emergent phenomena and they result from the catalytic role of genes, \citep{tyson}.

Another consequence of the modeling approach presented here is that positional information in developmental processes can be described by non-homogeneity in the spatial distribution of regulators, and not necessarily associated with other physical processes of transport and diffusion. 

Finally, we have validated a genetic regulatory network for the production of Hunchback and KNI during the 14th cleavage stage of the embryo of \textit{Drosophila}, and we have presented evidence that gap gene protein segments are out of equilibrium patterns. The genetic regulatory network of Figure~\ref{fig1} describes well the gap gene protein concentration of HB in the embryo length region $[5\%, 80\%]$, as well as the gap gene protein concentration of KNI in the embryo lenght region $[20\%, 100\%]$.

\section*{Acknowledgments}
This work has been supported  by European project GENNETEC, FP6 STREP IST 034952.


\begin{thebibliography}{}

\bibitem[Alves and Dil\~{a}o, 2005]{AD05} 
Alves,F. and Dil\~{a}o,R. (2005) 
A simple framework to describe the regulation of gene expression in prokaryotes. 
{\it C. R. Biologies}, {\bf 328}, 429-444.

\bibitem[Alves and Dil\~{a}o, 2006]{AD06} 
Alves,F. and Dil\~{a}o,R. (2006) 
Modeling segmental patterning in Drosophila: Maternal and gap genes. 
{\it J. The. Bio.}, {\bf 241}, 342-359.

\bibitem[Alves and Dil\~{a}o, 2006]{AD06b}
Alves,F. and Dil\~{a}o,R. (2006) 
A software tool to model genetic regulatory networks: applications to segmental patterning in Drosophila.
\textit{BIOMAT 2005, International Symposium on Mathematical and Computational Biology.}, 
Mondaini,R.P. and Dil\~{a}o,R. pp. 71-88, World Scientific, Singapore, 2006.

\bibitem[Dil\~{a}o \textit{et al.}, 2009]{Dilao}
Dil\~ao,R., Muraro,D., Nicolau,M. and  Schoenauer,M. (2009) Validation of a morphogenesis model of Drosophila early development by a multi-objective evolutionary optimization algorithm. In Pizzuti, C.,   Ritchie,M.D. and Giacobini,M. (Eds.): EvoBIO 2009, \textit{Lecture Notes in Computer Science} \textbf{5483}, 176–-190. 

\bibitem[Dil\~{a}o and Muraro, 2009]{DM09} 
Dil\~{a}o,R. and  Muraro,D. (2009) 
mRNA diffusion explains protein gradients in Drosophila early development. Pre-print.

\bibitem[Dil\~{a}o and Muraro, 2009b]{Dilao2} 
Dil\~{a}o,R. and  Muraro,D. (2009) 
Calibration of a genetic network model describing the production of gap gene proteins in Drosophila. Pre-print.

\bibitem[Driever and N\"{u}sslein-Volhard, 1988]{Driever}  Driever,W. and N\"{u}sslein-Volhard,C. (1988) A gradient of bicoid protein in \textit{Drosophila} embryos. \textit{Cell} 54, 83--93.

\bibitem[Jong, 2002]{jong} Jong,H.D. (2002) Modelling and simulations of genetic regulatory systems: a literature review. {\sl J.  Comput. Biol.} {\bf 9}, 67-103.

\bibitem[Jacob and Monod, 1961]{JM61} 
Jacob,F. and Monod,J. (1961) 
Genetic regulatory mechanisms in the synthesis of proteins.
{\it J. Mol. Biol.}, {\bf 3}, 318-356.

\bibitem[Klipp \textit{et al.}, 2005]{klipp} 
Klipp,E., Herwig,R., Kowald,A., Wierling,C. and   Lehrach,H. (2005)  Systems Biology in Practice.
Wiley-VCH, Weinheim.

\bibitem[N\"{u}sslein-Volhard, 1992]{Nuss} 
N\"{u}sslein-Volhard,C. (1992) Gradients that organize embryo development. \textit{Scientific American} 275(2), 54--61.

\bibitem[Pisarev  \textit{et al.}, 2009]{Pisarev} Pisarev,A., Poustelnikova,E., Samsonova,M. and Reinitz,J. (2009) FlyEx, the quantitative atlas on segmentation
gene expression at cellular resolution. \textit{Nucleic Acids Research} \textbf{37}, D560-–D566. 

\bibitem[Poustelnikova  \textit{et al.}, 2004]{Poustelnikova} Poustelnikova,E., Pisarev,A., Blagov,M., Samsonova,M. and  Reinitz,J. (2004) A database for management of gene expression data in situ. \textit{Bioinformatics} \textbf{20}, 2212--2221. 

\bibitem[Tyson, 2007]{tyson} Tyson,J.J. (2007) Bringing cartoons to life. \textit{Nature} \textbf{445}, 823. 

\bibitem[van Kampen, 1992]{kampen} 
van Kampen,N.G. (1992) Stochastic Processes in Physics and Chemistry.
North-Holland, Amsterdam.

\end{thebibliography}
\end{document}